\documentclass[twocolumn,showpacs,preprintnumbers,amsmath,amssymb]{revtex4}
\usepackage{graphicx}% Include figure files
\usepackage{dcolumn}% Align table columns on decimal point
\usepackage{bm}% bold math

\begin{document}

\preprint{APS/123-QED}

\title{On a Class of Spatial Discretizations of Equations of the
Nonlinear Schr{\"o}dinger  Type}

\author{P. G. Kevrekidis$^1$, S. V. Dmitriev$^2$, and
A. A. Sukhorukov$^3$%, and Yu. S. Kivshar$^3$
}
% \altaffiliation[Also at ]{Physics Department, XYZ University.}
\affiliation{ $^1$ Department of Mathematics and Statistics,
University of Massachusetts Lederle Graduate Research Tower,
Amherst, MA 01003-4515, USA \\
$^2$ Institute of Industrial Science, University of Tokyo, 4-6-1
Komaba, Meguro-ku, Tokyo 153-8505, Japan \\
$^3$Nonlinear Physics
Centre, Research School of Physical Sciences and Engineering, \\
Australian National University, Canberra ACT 0200, Australia }

\date{\today}

\begin{abstract}
We demonstrate the systematic derivation of a class of
discretizations of nonlinear Schr{\"o}dinger (NLS) equations for
general polynomial nonlinearity whose stationary solutions can be
found from a reduced two-point algebraic condition. We then focus
on the cubic problem and illustrate how our class of models
compares with the well-known discretizations such as the standard
discrete NLS equation, or the integrable variant thereof. We also
discuss the conservation laws of the derived generalizations of
the cubic case, such as the lattice momentum or mass and the
connection with their corresponding continuum siblings. (This
manuscript was submitted for publication on October 14, 2005.)
\end{abstract}


\maketitle

\section{Introduction}

In the past few years, the role of spatial discreteness in
lattice systems described by differential-difference equations
has been increasingly recognized \cite{reviews}. In these settings,
the spatial variables are discrete, while the
evolution variable is continuum. Relevant applications
are continuously arising in
rather diverse physical contexts such as the spatial
dynamics of optical beams in coupled waveguide arrays in
nonlinear optics \cite{reviews1},
the temporal evolution of Bose-Einstein condensates (BECs)
in optical lattices
in soft-condensed matter physics \cite{reviews2},
the DNA double strand in biophysics \cite{reviews3}, and so on.

In the examples that stem from physical applications, the form of
the discrete model is dictated by the underlying physics, and
typically that form is the discrete nonlinear Schr{\"o}dinger
(DNLS) equation \cite{dnls}. However, there are also exceptions to
this rule, where the nature of the nonlinearity \cite{hadz} or of
the dispersion \cite{efrem} or both \cite{johan} imposes
variations of this ubiquitous model. Another motivation to study
such modified DNLS models is a more mathematical one, namely the
aim of identifying models with good mathematical properties (e.g.
exact solutions, additional symmetries or, possibly, complete
integrability). Such a program was initiated by the derivation of
an integrable analog of the DNLS equation, namely the so-called
Ablowitz-Ladik model \cite{AL}. This was later used for
computational studies of the NLS equation \cite{schober}, as well
as implemented as a good starting point for developing
perturbation theoretic approaches to the DNLS limit, in order to
examine the existence and stability of its solutions
\cite{kapitula}.

We should note here that a similar, mathematically-minded
program of discretizations has been evolving for Klein-Gordon (KG)
type lattices. For instance, integrable discretizations of equations
such as the sine-Gordon model have been obtained
\cite{orfanidis,pilloni}; there have also been attempts to
systematically discretize preserving symmetries of the underlying
continuum model, such as, especially, a discrete analog of translational
invariance. It has often been noted that ``standard'' (e.g.
centered-difference schemes for Laplacian type operators) discretizations
strongly violate translational invariance, leading stable and
unstable steady states (typically centered on-site and inter-site
between two lattice nodes). This not only imposes undesirable
mathematical properties \cite{bender}, but also modifies the
underlying phenomenology in comparison to the continuum model \cite{peyrard}.
In that view, discretizations that preserve an effective translational
invariance by allowing the center of a stationary state to be a free
parameter (rather than to be fixed on- or half-way between two lattice
sites) have been sought. In this way, such discretizations also
avoid ``energy barriers'' (so called Peierls-Nabarro barriers)
between on-site and inter-site states. Such discrete models
have been constructed in the KG case, based on a discretization
of the energy, using the Bogomol'nyi approach \cite{speight}, as well as a
discretization of the equation of motion ensuring the persistence
of a discrete momentum conservation law \cite{PhysicaD,JPA}. These
classes of models were subsequently tested for the potential bearing
of travelling wave solutions, using the technology based on the calculation
of the Stokes constant \cite{pelin}; this led to the conjecture
that such models may possess isolated, exact, travelling lattice solutions.
It is, finally, worthwhile to note that while motivated by their
mathematical properties, such models may also bear physical
relevance as is indicated e.g. in the very recent preprint of
\cite{speight2} for the sine-Gordon case.

While this technology has been well developed for the single
(i.e., scalar) field case of KG lattices, such considerations do
not seem to have been applied to NLS type lattices, to the best of
our knowledge. The present manuscript aims to partially fill this
gap, by presenting a systematic methodology for deriving
discretizations of polynomial nonlinearity partial differential
equations (PDEs) of the NLS type. The main novel feature of these
discrete models is that, contrary to what is the case for the
standard DNLS equation, they preserve a discrete analog of the
momentum conservation law. In fact, we show that in the cubic
case, they are natural homotopic generalizations of the integrable
NLS discretization of \cite{AL}.

Our presentation will be structured as follows. In section II,
we present the general setup of the continuum and discrete models
of the present work. In section III, we illustrate the auxiliary
problem that aids us to construct the desired discretizations in
section IV. In section V, we study some of the
conservation laws of the obtained models, while in section VI,
we discuss their solitonic properties. Finally, in section VII,
we briefly summarize our findings and present our conclusions.

\section{Setup}

We present our methodology for the generalized NLS equation of the form
\begin{eqnarray}
\psi_t+\frac{1}{2} \psi_{xx}+G^{\prime}(|\psi|^2)\psi =0,
\label{NLSE}
\end{eqnarray}
where $\psi(x,t)$ is a complex function of two real variables;
$G(\xi)$ is a real function of its argument and
$G^{\prime}(\xi)=dG/d\xi$.

We introduce the lattice $x_n=nh$, where $h$ is the lattice
spacing and $n=0,\pm 1,\pm 2,...$  We also introduce the
following shorthand notations
\begin{eqnarray}
\psi_{n-1}=\psi_{-},\,\,\,\,\,{\rm
and}\,\,\,\,\,\psi_{n+1}=\psi_{+}, \label{Notations}
\end{eqnarray}
and will focus only on discretizations that involve such nearest
neighbor sites.

Our more specific aim will be
to construct the discrete analogues of Eq. (\ref{NLSE})
of the form:
\begin{eqnarray}
i \dot{\psi}_n +
r(\psi_{-},\psi_{-}^{\star},\psi_n,\psi_n^{\star},\psi_{+},\psi_{+}^{\star})=0,
\label{deq1}
\end{eqnarray}
such that the ansatz
\begin{eqnarray}
\psi_n(t)=f_n e^{i\omega t}, \label{AnsatzDiscrete}
\end{eqnarray}
reduces Eq. (\ref{deq1}) to the three-point discrete problem of
the form
\begin{eqnarray}
-\omega f_n + R(f_{-},f_{n},f_{+}) = 0, \label{Threepoint}
\end{eqnarray}
whose solution can be found from a reduced two-point discrete
problem $u(f_{-},f_{n}) = 0$. Such a selection will entail a
mono-parametric freedom for the resulting algebraic problem
leading to stationary state solutions.
This will, in turn, be  responsible for the effective translational
invariance in what follows.

\section{Auxiliary problem}

Firstly, we formulate an auxiliary problem. Seeking stationary
solutions of Eq. (\ref{NLSE}) in the form
\begin{eqnarray}
\psi(x,t)=f(x)e^{i\omega t}, \label{Ansatz}
\end{eqnarray}
we reduce it to an ordinary differential equation (ODE)
for the real function $f(x)$,
\begin{eqnarray}
D(x)\equiv f^{\prime \prime} - 2\omega f + 2fG^{\prime}(f^2) = 0,
\label{KGstatic}
\end{eqnarray}
having the first integral
\begin{eqnarray}
u(x)\equiv \left(f^{\prime}\right)^2 - 2\omega f^2 + 2G(f^2) = 0.
\label{KGstaticFI}
\end{eqnarray}

We then identify discretizations of Eq. (\ref{KGstatic}) of the form
\begin{eqnarray}
D(f_{-},f_n,f_{+}) = 0, \label{KGstaticDiscrete}
\end{eqnarray}
such that solutions to the three-point discrete Eq.
(\ref{KGstaticDiscrete}) can be found from a reduced two-point
problem
\begin{eqnarray}
u(f_{-},f_n)\equiv \frac{1}{h^2}\left(f_n-f_{-}\right)^2 \nonumber
\\- 2\omega f_{-}f_n + 2G(f_{-}^2,f_n^2) = 0,
\label{ReducedDiscrete}
\end{eqnarray}
which is a discrete version of Eq. (\ref{KGstaticFI}), assuming
that $G(f_{-}^2,f_n^2)$ reduces to $G(f^2)$ in the continuum limit
($h\rightarrow 0$).

Taking into account that Eq. (\ref{KGstatic}) is the static
Klein-Gordon equation with the potential
\begin{eqnarray}
V(f)=\omega f^2 - G(f^2), \label{Potential}
\end{eqnarray}
a wide class of discretizations solving the auxiliary problem has
been offered in the very recent work of \cite{JPA}.

For example, discretizing the left-hand side of the identity
$(1/2)du/df=D(x)$, we obtain the discrete version of Eq.
(\ref{KGstatic}),
\begin{eqnarray}
D_1(f_{-},f_n,f_{+})\equiv
\frac{u(f_n,f_{+})-u(f_{-},f_n)}{f_{+}-f_{-}}=0. \label{D1}
\end{eqnarray}
Formally, $D_1(f_{-},f_n,f_{+})=0$ is a three-point problem but,
clearly, its solutions can be found from the two-point problem
$u(f_{-},f_n)=0$ and thus, the auxiliary problem is solved. We note,
in passing,
that this type of argument was first proposed in
\cite{PhysicaD}.

\section{Main problem}

Coming back to our main problem of finding special discretizations
for Eq. (\ref{NLSE}), we should remark that among the solutions to the
auxiliary problem we should select the ones which can be rewritten
in terms of $\psi_n$ and $\psi_n^{\star}$ in the desired form of
Eq. (\ref{deq1}). This can be done when $D_1$ given by Eq.
(\ref{D1}) is written in a non-singular form (i.e., if the
denominator cancels with an appropriate factoring of the
numerator).  This always occurs if $G(\xi)$ is polynomial and if $u(f_{-},f_n)$
possesses the symmetry
\begin{eqnarray}
u(f_{-},f_n)=u(f_n,f_{-}). \label{Symmetry}
\end{eqnarray}

We thus focus on $G(\xi)$ in the form of Taylor expansion,
\begin{eqnarray}
G(|\psi|^2)=\sum_{k=1}^{\infty}a_k\left(|\psi|^2\right)^{k},
\label{Gpolynomial}
\end{eqnarray}
with real coefficients $a_k$; retaining first four terms of the
expansion, we write
\begin{eqnarray}
G(f_{-}^2,f_n^2)=\frac{a_1}{2}
\left(f_{-}^2+f_n^2\right)\nonumber \\
+ a_2\left[\frac{\alpha}{2}\left(f_{-}^4+f_n^4\right)
+(1-\alpha)f_{-}^2f_n^2\right] \nonumber \\
+ \frac{a_3}{2}\left[\beta\left(f_{-}^6+f_n^6\right)
+ (1-\beta) f_{-}^2f_n^2(f_{-}^2+f_n^2) \right] \nonumber \\
+ a_4\Big[\frac{\gamma}{2}\left(f_{-}^8+f_n^8\right)
+\frac{\delta}{2}f_{-}^2f_n^2(f_{-}^4+f_n^4) \nonumber \\
+ (1-\gamma-\delta)f_{-}^4f_n^4\Big], \label{GSymmetric}
\end{eqnarray}
where $\alpha,\beta,\gamma$, and $\delta$ are free parameters.
The symmetry condition of Eq. (\ref{Symmetry}) is satisfied for
$u(f_{-},f_n)$ given by (\ref{ReducedDiscrete}) since in Eq.
(\ref{GSymmetric}) we have $G(f_{-}^2,f_n^2)=G(f_n^2,f_{-}^2)$.

Equation (\ref{D1}) with $u(f_{-},f_n)$ given by Eq.
(\ref{ReducedDiscrete}) and $G(f_{-}^2,f_n^2)$ given by Eq.
(\ref{GSymmetric}) assumes the following form
\begin{eqnarray}
-\omega f_n + \frac{1}{2h^2}\left( f_{-} - 2f_n + f_{+}
\right) \nonumber \\
+\left( f_{-} + f_{+} \right)R(f_{-}^2,f_{n}^2,f_{+}^2) =0\,,
\label{Preliminary}
\end{eqnarray}
where
\begin{eqnarray}
R(f_{-}^2,f_{n}^2,f_{+}^2)=\frac{a_1}{2} +a_2 \left[
\frac{\alpha}{2}
\left( f_{-}^2+f_{+}^2\right)+(1-\alpha)f_n^2\right] \nonumber \\
+ \frac{a_3}{2}\beta\left( f_{-}^4 +f_{-}^2f_{+}^2 +
f_{+}^4\right) \nonumber \\
+ \frac{a_3}{2}(1-\beta)f_{n}^2\left( f_{-}^2 +f_{n}^2 +
f_{+}^2\right) \nonumber \\
+ a_4 \left( f_{-}^2 + f_{+}^2 \right) \left[ \frac{\gamma}{2}
\left( f_{-}^4 + f_{+}^4 \right) + (1-\gamma-\delta)f_n^4 \right] \nonumber \\
+ a_4 \frac{\delta}{2}f_n^2\left( f_{-}^4 + f_{n}^4 + f_{+}^4+
f_{-}^2 f_{+}^2 \right) .\,\,\, \label{RealMult}
\end{eqnarray}

One can conclude that the  discretization of NLS equation in the
form:
\begin{eqnarray}
i \dot{\psi}_n + \frac{1}{2h^2}\left( \psi_{-} - 2\psi_n +
\psi_{+} \right) \nonumber \\
+\left( \psi_{-} + \psi_{+}
\right)R(|\psi_{-}|^2,|\psi_{n}|^2,|\psi_{+}|^2)=0\,,
\label{DNLSE}
\end{eqnarray}
with $R$ given by the expression of (\ref{RealMult}) satisfies the
generalized equation (\ref{NLSE}) with $G(\xi)$ given by Eq.
(\ref{Gpolynomial}). However, as per the construction above,
additionally, the corresponding discrete equation for the
stationary solutions of the form of Eq. (\ref{AnsatzDiscrete}) is
the three-point problem of Eq. (\ref{Preliminary}), whose solution
can be found through the two-point reduction of Eq.
(\ref{ReducedDiscrete}) where $G(f_{-}^2,f_n^2)$ is given by Eq.
(\ref{GSymmetric}).

It is interesting to note that the ``standard'' DNLS equation
\begin{eqnarray}
i\dot{\psi}_n + \frac{1}{2h^2}( \psi_{-} - 2\psi_n +
\psi_{+})+|\psi_n|^2\psi_n=0,\label{DNLSEclassic}
\end{eqnarray}
does not belong to the above class and more generally does not
share the reduction property used above. Instead, and focusing
only on the cubic Kerr nonlinearity (where $G^{\prime}$ is linear
in its argument), we obtain from Eq. (\ref{DNLSE}) and Eq.
(\ref{RealMult})
\begin{eqnarray}
i \dot{\psi}_n + \frac{1}{2h^2}\left( \psi_{-} - 2\psi_n +
\psi_{+} \right) %\nonumber \\
+\frac{1}{2}\left( \psi_{-} + \psi_{+} \right) \nonumber \\
\times \left[ \frac{\alpha}{2} \left(
|\psi_{-}|^2+|\psi_{+}|^2\right)+(1-\alpha)|\psi_n|^2\right] =0.
\label{DNLSEAL}
\end{eqnarray}
Notice that the integrable discretization of \cite{AL} is obtained
from this approach as the special case of $\alpha=0$. For $\alpha
\neq 0$, this model can be regarded as a Salerno-type model
\cite{Salerno}, i.e., a homotopic continuation including the
integrable limit and reducing to NLS equation in the continuum
limit.

Expression (\ref{GSymmetric}) contains only the terms with even
powers of $f_{-}$ and $f_n$. However, it is possible to construct
the terms of desired symmetry involving odd powers \cite{JPA}. For
example, for Kerr nonlinearity one can take
\begin{eqnarray}
G(f_{-},f_n)=\frac{1}{4}f_{-}f_n\left(f_{-}^2+f_n^2\right),
\label{A1}
\end{eqnarray}
and obtain from Eq. (\ref{ReducedDiscrete}) and Eq. (\ref{D1})
\begin{eqnarray}
-\omega f_n + \frac{1}{2h^2}\left( f_{-} - 2f_n + f_{+}
\right) \nonumber \\
+\frac{f_n^3}{4}+\frac{f_n}{4}\left( f_{-}^2 +f_{-}f_{+}+ f_{+}^2
\right) =0\,, \label{A2}
\end{eqnarray}
for which the following discretization can be obtained
\begin{eqnarray}
i \dot{\psi}_n + \frac{1}{2h^2}\left( \psi_{-} - 2\psi_n +
\psi_{+} \right) %\nonumber \\
+\frac{1}{4}|\psi_n|^2\psi_n \nonumber \\
+\frac{1}{4}\left(
|\psi_{-}|^2+|\psi_{-}\psi_{+}|+|\psi_{+}|^2\right)\psi_n =0.
\label{A3}
\end{eqnarray}
The model of Eq. (\ref{A3}) has {\em on-site} cubic nonlinearity
modified through inter-site coupling, which makes it qualitatively
different from integrable system of \cite{AL}. Soliton
solutions to this model can be constructed from the quartic Eq.
(\ref{ReducedDiscrete}) with $G$ given by Eq. (\ref{A1}). Note
that DNLS equation with the quintic term possessing a structure similar to
the last term of Eq. (\ref{A3}) has been considered in
\cite{PhysicaD} [see Eq. (74) of that paper].

\section{Momentum and Mass Conservation Laws}

We now try to connect the above presented construction to the relevant
conservation laws of the resulting infinite-dimensional dynamical
system. More specifically, we examine the momentum conservation
law that, as argued above, is intimately related to the
translational invariance and the existence of mono-parametric
stationary solutions. For the DNLS model of Eq. (\ref{deq1}), we
consider the momentum defined as
\begin{eqnarray}
P=i \sum_{n=-\infty}^{\infty} \left( \psi_n \psi_{+}^{\star} -
\psi_n^{\star} \psi_{+} \right) \nonumber \\
\equiv i \sum_{n=-\infty}^{\infty} \psi_n \left( \psi_{+} -
\psi_{-} \right)^{\star}. \label{MomNLSE}
\end{eqnarray}

We now demand that the momentum be conserved, i.e., that
\begin{eqnarray}
\frac{dP}{dt}=0. \label{deq5}
\end{eqnarray}
Upon substitution of Eq. (\ref{MomNLSE}) and use of the equation
of motion Eq. (\ref{deq1}), $i \dot{\psi}_n + r_n=0$, for
$\dot{\psi}_n$, $\dot{\psi}_{+}^{\star}$ and
$\dot{\psi}_{-}^{\star}$, we obtain
\begin{eqnarray}
\sum_{n=-\infty}^{\infty} \left(r_{+}^{\star} \psi_n - r_{n}
\psi_{+}^{\star} \right) + \sum_{n=-\infty}^{\infty}\left(r_{+}
\psi_n^{\star} -r_n^{\star} \psi_{+} \right)=0. \label{deq5a}
\end{eqnarray}
In the last expression, if the first sum is zero, then the second
sum is also zero, being its complex conjugate. Thus, a sufficient
condition for the conservation of the momentum is
\begin{eqnarray}
\sum_{n=-\infty}^{\infty} \left(r_{n}^{\star} \psi_{-} - r_{n}
\psi_{+}^{\star} \right) = 0. \label{MomConsCond}
\end{eqnarray}
A direct consequence of the above is that the regular DNLS
equation, will not succeed in leading to conservation of momentum,
while it can easily be checked that the opposite is true for the
model of Eq. (\ref{DNLSE}) with $R$ given by Eq. (\ref{RealMult}).

The Eq. (\ref{DNLSE}) with $R$ given by Eq. (\ref{RealMult})
conserves momentum Eq. (\ref{MomNLSE}) but it does not conserve
the ``standard'' ($l^2$) norm
\begin{eqnarray}
N=\sum_n |\psi_n|^2. \label{Norm}
\end{eqnarray}
 In fact, it can be shown that unless $R$ is
``local'' (i.e., dependent on $|\psi_n|^2$ and not its neighbors,
as is e.g. the case for the integrable discretization of
\cite{AL}), there is no definition of $N=\sum_n F(|\psi_n|^2)$
which can be preserved under (\ref{DNLSE}). Instead Eq.
(\ref{DNLSE}) preserves the ``mass'' (norm) of the form:
\begin{eqnarray}
\tilde{N}=\sum_n \psi_n \left(\psi_{+}^{\star} + \psi_{-}^{\star}
\right), \label{newnorm}
\end{eqnarray}
which in the continuum limit retrieves the standard conservation
law of the $L^2$ norm. Notice, however, that the discretization of
Eq. (\ref{A3}) does conserve the standard norm.

For reasons of comparison that will become more transparent below, let us also
introduce an additional discretization that does {\it not} belong
to the family of Eq. (\ref{DNLSE}) presented above:
\begin{eqnarray}
i \dot{\psi}_n + \frac{1}{2h^2}\left( \psi_{-} - 2\psi_n +
\psi_{+} \right) %\nonumber \\
+\frac{1}{4}\left( \psi_{-} + \psi_{+} \right)|\psi_n|^2 \nonumber \\
+\frac{1}{4}\left( \psi_{-}^{\star} + \psi_{+}^{\star}
\right)\psi_n^2 =0. \label{DNLSE1}
\end{eqnarray}
The particular feature of this dynamical system is that it
conserves the $l^2$ norm but does {\it not} conserve the momentum
defined above. However, considering the standing wave ansatz of
Eq. (\ref{AnsatzDiscrete}) reduces Eq. (\ref{DNLSE1}) to
\begin{eqnarray}
-\omega f_n + \frac{1}{2h^2}\left( f_{-} - 2f_n + f_{+}
\right) %\nonumber \\
+\frac{1}{2}\left( f_{-} + f_{+} \right)f_{n}^2 =0,
\label{ParticularCase}
\end{eqnarray}
which is particular case of Eq. (\ref{Preliminary}) at $a_2=1/2$,
$a_1 = a_3 = a_4 = 0$ and $\alpha = 0$. Thus, solution of Eq.
(\ref{ParticularCase}) can be found from the two-point difference
equation
\begin{eqnarray}
u(f_{-},f_n)\equiv \frac{1}{h^2}\left(f_n-f_{-}\right)^2 %\nonumber \\
- 2\omega f_{-}f_n + f_{-}^2f_n^2 = 0,
\label{ReducedDiscreteParticular}
\end{eqnarray}
which is particular case of Eq. (\ref{ReducedDiscrete}). As a
result, the stationary states can be found by the solution of a
reduced two-point problem; namely, for any $f_{-}$ (or $f_{n}$) in
the range $[q_m,q_s]$, where $q_m=\sqrt{2\omega}$ and
$q_s=h^{-1}\sqrt{1-(1+\omega h^2)^{-2}}$, solving the binomial Eq.
(\ref{ReducedDiscreteParticular}), one can find $f_{n}$ (or
$f_{-}$), thus reconstructing the soliton solution for Eq.
(\ref{DNLSE1}) in the form of Eq. (\ref{AnsatzDiscrete}).
Quantities $q_m$ and $q_s$ are the amplitudes of solitons centered
between two lattice sites and on a lattice site, respectively.
Interestingly, $q_m$ is not a function of $h$. Even more
interestingly perhaps, such stationary solutions of Eq.
(\ref{DNLSE1}) are identical (see also below) to those of the
integrable equation [Eq. (\ref{DNLSEAL}) at $\alpha=0$] since
the two models share the same reduced two-point problem [Eq.
(\ref{ReducedDiscreteParticular})].

\section{Comparison of soliton solutions for different discretizations}

We now compare some properties of the classical DNLS model of Eq.
(\ref{DNLSEclassic}) (model $I$), the ``secular'' model
of Eq. (\ref{DNLSE1}) conserving the classical norm (model $II$),
and the Kerr-representative of the class of
models developed herein, given by Eq. (\ref{DNLSEAL}) (model $III$).

All three models share the same continuum limit, the integrable
NLS equation with Kerr nonlinearity,
\begin{eqnarray}
\psi_t+\frac{1}{2} \psi_{xx}+|\psi|^2\psi =0, \label{NLSEKerr}
\end{eqnarray}
and thus, in the regime of weak discreteness (small lattice spacing
$h$), their
soliton solutions of the form of Eq. (\ref{AnsatzDiscrete}) can be
expressed {\it approximately} as
\begin{eqnarray}
\psi_{n}(t)=\frac{q}{\cosh[q h (n-x_0) ]}\exp[-i(q^2/2)t],
\label{ApproxSOLITON}
\end{eqnarray}
where $q$ and $\omega=q^2/2$ are the soliton amplitude and
frequency, respectively.

The approximate solution of Eq. (\ref{ApproxSOLITON}) contains the
free parameter $x_0$ defining the soliton position. However, as
indicated above, in contrast to the NLS equation of Eq.
(\ref{NLSEKerr}), where $x_0$ can be chosen arbitrarily due
to translational invariance, the DNLS models {\it usually} have
stationary soliton solutions only for a discrete set of values of
$x_0$ (e.g. on-site and inter-site, as mentioned above). This is
true, for example, for the classical DNLS of model $I$ and for the
Salerno model \cite{Salerno}, among others. The models $II$ and
$III$, by construction, are among the members of a wider class of DNLS
equations proposed in this paper, where stationary soliton
solutions exist for any $x_0$, or, in other words, they can be
placed anywhere with respect to the lattice; otherwise put,
the Peierls-Nabarro potential is absent for stationary solutions
of these models.

An explicit formula, as is well-known \cite{dnls}, does not exist
for the solutions of model $I$. Such solutions can be obtained
numerically with the desired degree of accuracy, for two
particular cases of $x_0=0$ and $x_0=1/2$ (due to the integer
shift-invariance of the lattice, we now restrict ourselves to
$0\leq x_0 \leq 1$). To obtain the soliton of frequency $\omega$
centered on a lattice site, i.e., the one with $x_0=0$, we set
$f_0=\sqrt{2\omega}$ which is the soliton amplitude estimated from
Eq. (\ref{ApproxSOLITON}). Having $f_0$ and the symmetry property
$f_n=f_{-n}$ for $n>0$, we find successively $f_1$, then $f_2$ and
so on from the equation $-\omega f_n + (2h^2)^{-1}\left( f_{-} -
2f_n + f_{+} \right) + f_{n}^3 =0$, which is obtained by
substituting Eq. (\ref{AnsatzDiscrete}) into Eq.
(\ref{DNLSEclassic}). Since $f_0$ is not an exact value of the
soliton amplitude, the boundary conditions
$f_{-\infty}=f_{\infty}=0$ will not be satisfied. We find
numerically a correction to $f_0$ to satisfy the boundary
conditions thus completing the construction of the soliton
solution. For $x_0$ different from 0 or 1/2 it is impossible to
satisfy both boundary conditions simultaneously. The soliton with
$x_0=1/2$ can be constructed similarly using the symmetry property
$f_{-1}=f_0$ and $f_n=f_{-n-1}$ for $n>0$. Estimation of the soliton
amplitude from Eq. (\ref{ApproxSOLITON}) in this case is
$f_0=\sqrt{2\omega}/\cosh[(h/2)\sqrt{2\omega}]$. Alternatively,
one can straightforwardly use fixed point algorithms to obtain
such solutions as is summarized in \cite{dnls}.

For the models $II$ and $III$ the exact solutions of the form of
Eq. (\ref{AnsatzDiscrete}) can be found using the method developed
in section IV. More specifically, however, for model $II$, as is
expected from the discussion above and the coincidence of the
reduced two-point problem with that of the integrable discrete
model, an exact stationary solution can be obtained explicitly in
the form
\begin{eqnarray}
\psi_{n}(t)=\frac{1}{h}\frac{\sinh \mu}{\cosh[\mu (n-x_0)]}
\exp^{i\omega t}, \label{ALsoliton}
\end{eqnarray}
where $x_0$ is the parameter defining the soliton position and it
can obtain any value from $[0,1)$. The soliton frequency
$\omega=h^{-2}(1-\cosh\mu)$ and amplitude $q=h^{-2}\sinh^2\mu$ are
expressed in terms of the free parameter $\mu>0$. The surprising
feature of this secular discretization is that, despite the
absence of an explicit momentum conservation law, the stationary
solutions appear to enjoy an
effective invariance with respect to their center location.

Model $III$ has the solutions of the form of Eq.
(\ref{AnsatzDiscrete}) with $f_n$ derivable from the
two-point problem
\begin{eqnarray}
\frac{1}{h^2}(f_{-}-f_{n})^2-2\omega f_{-}f_{n} +
\frac{\alpha}{2}\left(
f_{-}^4 + f_{n}^4\right) \nonumber \\
+(1-\alpha)f_{-}^2 f_{n}^2=0. \label{ModelIII}
\end{eqnarray}
Equation (\ref{ModelIII}) is a particular case of Eq.
(\ref{ReducedDiscrete}) with $G$ given by Eq. (\ref{GSymmetric})
at $a_k=0$ for all $k$ except for $a_1=1/2$. The soliton can be
constructed by setting an arbitrary value for $f_{-}$ (or $f_{n}$)
in the range $[q_m,q_s]$ and finding $f_{n}$ (or $f_{-}$) from
the quartic Eq. (\ref{ModelIII}). Quantities $q_m$ and $q_s$ are
the amplitudes of solitons centered between two lattice sites and
on a lattice site, respectively. We have $q_m=\sqrt{2\omega}$,
which does not depend on $h$ and $\alpha$. For $f_{-}>q_s$ Eq.
(\ref{ModelIII}) does not have real solutions, i.e., $q_s$
corresponds to the magnitude of $f_{-}$ for which the two distinct
real roots of Eq. (\ref{ModelIII}) in $f_n$ merge into a multiple
root. The arbitrariness in the choice of initial value of $f_{-}$
(or $f_{n}$) implies the absence of the Peierls-Nabarro potential
and the possibility to place the soliton anywhere with respect to
the lattice.

\begin{figure}
\includegraphics{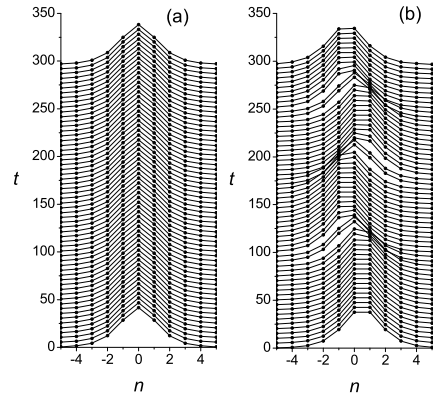}
\caption{The dynamical evolution of $|\psi_n|^2$ for exact
solitary wave solutions in the classical DNLS model $I$ centered
at (a) $x_0=0$ and (b) $x_0=1/2$. Soliton in (a) is stable while
in (b) it is unstable and, due to the presence of perturbations in
the form of round-off errors, it spontaneously starts to
alternate between two nearest inter-site
configurations passing through the stable on-site configuration.
Results for lattice spacing $h=0.4$ and soliton frequency
$\omega=1$.} \label{Figure1}
\end{figure}

\begin{figure}
\includegraphics{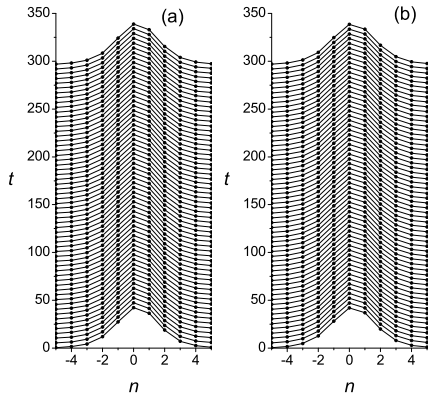}
\caption{The dynamical evolution of $|\psi_n|^2$ for exact
solitary wave solutions placed non-symmetrically with respect to
the lattice for (a) model $II$ and (b) model $III$ with
$\alpha=0.2$. In both cases the solitons do not radiate and they
do not move to a higher-symmetry position since the
Peierls-Nabarro potential is absent. Results for lattice spacing
$h=0.4$ and soliton frequency $\omega=1$.} \label{Figure2}
\end{figure}

The dynamical evolution of $|\psi_n|^2$ for exact solitary wave
solutions (constructed numerically as described above) in the
classical DNLS of model $I$ is shown in Fig. \ref{Figure1} where
we compare the solitons centered at (a) $x_0=0$ and (b) $x_0=1/2$.
The results are obtained by numerical integration of Eq.
(\ref{DNLSEclassic}) for lattice spacing $h=0.4$ and soliton
frequency $\omega=1$. The stationary solution in (a) is stable
while the inter-site centered one in (b) is unstable; due to
the presence of perturbations in the form of round-off errors, it
spontaneously starts to alternate
between the two nearest inter-site configurations passing through the
stable on-site configuration. The pulse of Fig. \ref{Figure1}(a)
does not radiate while that of Fig. \ref{Figure1}(b) does.

Similar computations have been carried out in models $II$ and
$III$, and are reported in Fig. \ref{Figure2}, also using $h=0.4$
and $\omega=1$. The solitary waves were found to be stable for any
$x_0$ both in model $II$ and model $III$ for negative and positive
$\alpha$ in a vicinity of $\alpha=0$. For example, in Fig.
\ref{Figure2} we show the results for (a) model $II$ and (b) model
$III$ with $\alpha=0.2$ and initial profiles placed
non-symmetrically with respect to the lattice. In both cases the
solitons do not radiate and they do not move to a higher-symmetry
position.

\section{Conclusions}

We have described a general and systematic method of constructing
spatial discretizations of NLS-type models, whose stationary
soliton solutions can be obtained from a two-point difference
problem. In this setting, finding stationary solutions becomes
tantamount to solving simple nonlinear algebraic equations. We
have also illustrated the connections of the resulting models with
the integrable discretization of the NLS equation, of which they
are a natural generalization for cubic nonlinearities (our
construction was given for arbitrary polynomial
nonlinearities of a particular parity);
furthermore, the differences of such models from
the standard discretization of the NLS equation often encountered
in physical applications have been highlighted, both in terms of
the relevant dynamical (solitonic) behavior as well as in terms of
the underlying conservation laws present in the various models.

It would be particularly interesting to further examine such
discretizations and their features, such as the stability of
their solutions \cite{dnls}, and their travelling wave properties
\cite{pelin,floria,pelin2} and potential integrability of
special members within these families. Such studies are currently
in progress and will be reported in future publications.

\section*{Acknowledgements}

We would like to acknowledge a number useful discussions with Yu. S.
Kivshar.  SVD wishes to thank the warm hospitality of the
Nonlinear Physics Centre at the Australian National University.
PGK gratefully acknowledges the support of NSF-DMS-0204585,
NSF-DMS-0505063 and NSF-CAREER.

%\end{multicols}

\end{document}